# Mode-Dependent Loss and Gain: Statistics and Effect on Mode-Division Multiplexing

**Keang-Po Ho[1,*] and Joseph M. Kahn[2]**

[1]*Silicon Image, Sunnyvale, CA 94085, USA.*
[2]*Department of Electrical Engineering, Stanford University, Stanford, CA 94305, USA*
[*]*kpho@ieee.org*

**Abstract:** In multimode fiber transmission systems, mode-dependent loss and gain (collectively referred to as MDL) pose fundamental performance limitations. In the regime of strong mode coupling, the statistics of MDL (expressed in decibels or log power gain units) can be described by the eigenvalue distribution of zero-trace Gaussian unitary ensemble in the small-MDL region that is expected to be of interest for practical long-haul transmission. Information-theoretic channel capacities of mode-division-multiplexed systems in the presence of MDL are studied, including average and outage capacities, with and without channel state information.

**OCIS codes:** (060.2400) Fiber properties; (060.2330) Fiber optics communications; (000.5490) Probability theory, stochastic processes, and statistics

---

**References and Links**

1. IEEE 802.3 Standard, Carrier Sense Multiple Access with Collision Detection (CSMA/CD) Access Method and Physical Layer Specifications, 2008.
2. A. F. Benner, M. Ignatowski, J. A. Kash, D. M. Kuchta, and M. B. Ritter, "Exploitation of optical interconnects in future server architectures," IBM J. Res. & Dev. **49**, 755-775 (2005).
3. Y. Koike and S. Takahashi, "Plastic optical fibers: technologies and communication links," in *Optical Fiber Telecommunications VB: Systems and Networks*, I. P. Kaminow, T. Li and A. E. Willner eds. (Elsevier Academic, 2008).
4. D. Gloge, "Optical power flow in multimode fibers," Bell System Tech. J. **51**, 1767-1780 (1972).
5. R. Olshansky, "Mode-coupling effects in graded-index optical fibers," App. Opt. **14**, 935-945 (1975).
6. L. Raddatz, I. H. White, D. G. Cunningham, and M. C. Nowell, "An experimental and theoretical study of the offset launch technique for the enhancement of the bandwidth of multimode fiber links," J. Lightwave Technol. **16**, 324-331 (1998).
7. K.-P. Ho and J. M. Kahn, "Statistics of group delays in multimode fiber with strong mode coupling," submitted to J. Lightwave Technol., http://arxiv.org/abs/1104.4527
8. M. L. Mehta, Random Matrices, 3rd ed. (Elsevier Academic, 2004).
9. B. Rosinski, J. W. D. Chi, P. Grosso, and J. Le Bihan, "Multichannel transmission of a multicore fiber coupled with vertical-cavity surface-emitting lasers," J. Lightwave Technol. **17**, 807-810 (1999).
10. B. Zhu, T. F. Taunay, M. F. Yan, J. M. Fini, M. Fishteyn, E. M. Monberg, and F. V. Dimarcello, "Seven-core multicore fiber transmissions for passive optical network," Opt. Express **18**, 11117-11122 (2010) http://www.opticsinfobase.org/abstract.cfm?URI=oe-18-11-11117.
11. H. R. Stuart, "Dispersive multiplexing in multimode optical fiber," Science **289**, 281-283 (2000).
12. A. Tarighat, R. C. J. Hsu, A. Shah, A. H. Sayed, and B. Jalali, "Fundamentals and challenges of optical multiple-input multiple-output multimode fiber links," IEEE Commun. Mag. (5), 57-63 (2007).
13. A. R. Shah, R. C. J. Hsu, A. Tarighat, A. H. Sayed, and B. Jalali, "Coherent optical MIMO (COMIMO)," J. Lightwave Technol. **23**, 2410-2419 (2005).
14. R. C. J. Hsu, A. Tarighat, A. Shah, A. H. Sayed, and B. Jalali, "Capacity enhancement in coherent optical MIMO (COMIMO) multimode fiber links," IEEE Commun. Lett. **10**, 1089-7798 (2006).
15. M. Nazarathy and A. Agmon, "Coherent transmission direct detection MIMO over short-range optical interconnects and passive optical networks," J. Lightwave Technol. **26**, 2037–2045 (2008).
16. H. Bülow, "Coherent multichannel transmission over multimode-fiber and related signal processing," Proc. of OSA Topical Meeting on Access Networks and In-House Communications, Karlsruhe, Germany, June 21-24, 2010, paper AThB1.
17. A. Li, A. Al Amin, X. Chen, and W. Shieh, "Reception of mode and polarization multiplexed 107-Gb/s CO-OFDM signal over a two-mode fiber," in OFC '11, paper PDPB8.

18. R. Ryf, S. Randel, A. H. Gnuack, C. Bolle, R.-J. Essiambre, P. Winzer, D. W. Peckham, A. McCurdy, and R. Lingle, "Space-division multiplexing over 10 km of three-mode fiber using coherent 6 × 6 MIMO processing," in OFC '11, paper PDPB10.
19. M. Salsi, C. Koebele, D. Sperti, P. Tran, P. Brindel, H. Margoyan, S. Bigo, A. Boutin, F. Verluise, P. Sillard, M. Bigot-Astruc, L. Provost, F. Cerou, and G. Charlet, "Transmission at 2 × 100 Gb/s, over two modes of 40 km-long prototype few-mode fiber, using LCOS based mode multiplexer and demultiplexer," in OFC '11, paper PDPB9.
20. Z. Tong, Q. Yang, Y. Ma, and W. Shieh, "21.4 Gbit/s transmission over 200 km multimode fiber using coherent optical OFDM", Electron. Lett. **44**, 1373-1374 (2008).
21. W. Shieh, H. Bao, and Y. Tang, "Coherent optical OFDM: theory and design," Opt. Express **16**, 841-859 (2008) http://www.opticsinfobase.org/abstract.cfm?URI=oe-16-2-841.
22. J. Shentu, K. Pantu, and J. Armstrong, "Effects of phase noise on performance of OFDM systems using an ICI cancellation scheme," IEEE Trans. Broadcasting **49**, 221-224 (2003).
23. S. Wu and Y. Bar-Ness, "OFDM systems in the presence of phase noise: consequences and solutions," IEEE Trans. Commun. **52**, 1988-1996 (2004).
24. W. Shieh and K.-P. Ho, "Equalization-enhanced phase noise for coherent-detection systems using electronic digital signal processing," Opt. Express **16**, 15718-15727 (2008) http://www.opticsinfobase.org/abstract.cfm?URI=oe-16-20-15718.
25. C. Xie, "WDM coherent PDM-QPSK systems with and without inline optical dispersion compensation," Opt. Express **17**, 4815-4823 (2009) http://www.opticsinfobase.org/abstract.cfm?URI=oe-17-6-4815.
26. K.-P. Ho, A. P. T. Lau, and W. Shieh, "Equalization-enhanced phase noise induced timing jitter," Opt. Lett. **36**, 585-587 (2011) http://www.opticsinfobase.org/abstract.cfm?URI=ol-36-4-585.
27. P. J. Winzer and G. J. Foschini, "Outage calculations for spatially multiplexed fiber links," in OFC '11, paper OThO5.
28. J. P. Gordon and H. Kogelnik, "PMD fundamentals: polarization-mode dispersion in optical fibers," Proc. Natl. Acad. Sci. **97**, 4541-4550 (2000).
29. H. Kogelnik, R. M. Jopson, and L. E. Nelson, "Polarization-mode dispersion" in *Optical Fiber Telecommunications IVB: Systems and Impairments*, I. Kaminow and T. Li, eds, (Academic Press, 2002).
30. C. D. Poole and R. E. Wagner, "Phenomenological approach to polarization dispersion in long single-mode fibers," Electron. Lett. **22**, 1029-1030 (1986).
31. M. B. Shemirani, W. Mao, R. A Panicker, and J. M. Kahn, "Principal modes in graded-index multimode fiber in presence of spatial- and polarization-mode coupling," J. Lightwave Technol. **27**, 1248-1261 (2009).
32. A. M. Tulino and S. Verdú, *Random Matrix Theory and Wireless Communications*, (Now, 2004).
33. D. Tse and P. Viswanath, *Fundamentals of Wireless Communication*, (Cambridge Univ. Press, 2005).
34. X. Zhu and R. D. Murch, "Layered space-frequency equalization in a single-carrier MIMO system for frequency-selective channels," IEEE Trans. Wireless Commun. **3**, 701-708 (2004).
35. B. Vucetic and J. Yuan, *Space-Time Coding*, (Wiley, 2003).
36. V. Tarokh, H. Jafarkhani, and A. R. Calderbank, "Space-time block coding for wireless communications: performance results," IEEE J. Sel. Areas in Commun. **17**, (451-460) 1999.
37. R. Bellman, "Limit theorems for non-commutative operations I," Duke Math. J. **21** 491-500 (1954).
38. H. Furstenberg and H. Kesten, "Products of random matrices," Ann. Math. Stat. **31**, 457–469 (1960).
39. J. E. Cohen and C. M. Newman, "The stability of large random matrices and their products," Ann. Probab. **12**, 283-310 (1984)
40. A. Crisanti, G. Paladin, and A. Vulpiani, *Products of Random Matrices in Statistical Physics*, (Springer, 1993).
41. M. A. Berger, "Central limit theorem for products of random matrices," Trans. Am. Math. Soc. **285**, 777-803 (1984).
42. A. Mecozzi and M. Shtaif, "The statistics of polarization-dependent loss in optical communication systems," IEEE Photon. Technol. Lett. **14**, 313-315 (2002).
43. A. Glatarossa and L. Palmieri, "The exact statistics of polarization-dependent loss in fiber-optic links," IEEE Photon. Technol. Lett. **15**, 57-59, (2003).
44. D. Voiculescu, K. Dykema, and A. Nica, *Free Random Variables*, CRM Monograph Series, vol. 1, (American Mathematical Society, 1992).
45. A. Nica and R. Speicher, *Lectures on the Combinatorics of Free Probability*, London Mathematical Society Lecture Note Series, vol. 335, (Cambridge Univ. Press, 2006).
46. P. Lu, L. Chen, and X. Bao, "Statistical distribution of polarization dependent loss in the presence of polarization mode dispersion in single mode fibers," IEEE Photon. Technol. Lett., **13**, 451–453 (2001).
47. D. Voiculescu, "Limit laws for random matrices and free products," Invent. Math. **104**, 201-220 (1991).
48. T. Tao and V. H. Vu, "From the Littlewood-Offord problem to the circular law: Universality of the spectral distribution of random matrices," Bull. Am. Math. Soc. **46**, 337–396 (2009).
49. A. Papoulis, *Probability, Random Variables, and Stochastic Processes,* 2$^{nd}$ ed. (McGraw-Hill, 1984).
50. T. S. Rapport, *Wireless Communications: Principles & Practice*, (Prentice Hall, 1996).
51. K.-P. Ho, "Statistical properties of stimulated Raman crosstalk in WDM systems," J. Lightwave Technol. **18**, 915-921 (2000)

52. K.-P. Ho, "Central limits for the products of free random variables," http://arxiv.org/abs/1101.5220.
53. H. Bercovici and V. Pata, "Limit laws for products of free and independent random variables," Studia Math. **141**, 43–52 (2000)
54. G. Chistyakov and F. Götze, "Limit theorems in free probability theory. II," Central Eur. J. Math. **6**, 87–117 (2008).
55. V. Kargin, "The norm of products of free random variables," Probab. Theory Relat. Fields **139**, 397–413 (2007).
56. E. Wigner, "Characteristic vectors of bordered matrices with infinite dimensions," Ann. Math. **62**, 548-564 (1955).
57. H. G. Golub and C. F. van Loan, *Matrix Computations*, 3$^{rd}$ ed. (Johns Hopkins, 1996).
58. I. Dumitriu and A. Edelman, "Matrix models for beta ensembles," J. Math. Phys. **43**, 5830-5847(2002).
59. A. S. Hedayat, N. J. A. Sloane, and J. Stufken, *Orthogonal Arrays: Theory and Applications*, (Springer, 1999).

## 1. Introduction

Multimode fiber (MMF) has been employed traditionally in short-distance systems, including local-area networks and data-center interconnects [1-3]. In such applications, MMF is often favored over single-mode fiber (SMF) because of relaxed connector alignment tolerances and reduced transceiver component costs. MMF supports propagation of multiple spatial modes having different group delays and potentially different losses. Manufacturing variations, bends, mechanical stresses, thermal gradients and other effects cause coupling between different modes. Even if a signal is launched into a single spatial mode, it tends to couple into multiple spatial modes with different group delays. This modal dispersion [4-6] causes intersymbol interference (ISI), and is the primary factor limiting bandwidth-distance products in short-distance MMF systems using direct detection. Mode-dependent loss and gain (collectively referred to as MDL), are typically less important than modal dispersion in short-distance systems.

The characteristics of a MMF, in particular, the modal group delay profile and loss profile, vary along the length of a fiber, and can be considered constant only over a characteristic correlation length. When a fiber is much longer than the correlation length, it can be modeled as a concatenation of numerous sections with independent characteristics. We refer to this as the strong-coupling regime. In a recent paper [7], we studied modal dispersion in MMF in the strong-coupling regime, showing that it is statistically equivalent to a zero-trace random Gaussian Hermitian matrix, or a Gaussian unitary ensemble [8], with eigenvalues corresponding to modal group delays. In the strong-coupling limit, the modal group delays follow a limiting distribution that depends only on a single parameter and the number of modes [7].

In recent works, data rates have been increased by spatially multiplexing several parallel data streams in one fiber. This can be achieved using multi-core fiber [9-10], or by mode-division multiplexing (MDM) in MMF [11-19]. Long-haul systems using MDM in MMF are expected to use coherent detection. When using coherent detection, large modal dispersion may necessitate complex signal processing, but it does not fundamentally limit performance or channel capacity. For example, systems using orthogonal frequency-division multiplexing (OFDM) [20-21] can avoid ISI given sufficiently long cyclic prefix and narrow subcarrier spacing, but at the price of a reduced tolerance to fiber nonlinearity and phase noise, unless complex signal processing is used for inter-carrier interference cancellation [22-23]. Likewise, systems using single-carrier modulation can mitigate ISI given a sufficient number of equalizer taps, but at the price of equalization-enhanced phase noise [24-26].

Long-haul MDM systems will employ many inline optical components, including amplifiers and switches, which can introduce MDL. Unlike modal dispersion, MDL is fundamentally a performance-limiting factor. In the extreme case, MDL is equivalent to a reduction in the number of propagating modes, leading to a proportional decrease in data rate or channel capacity.

In this paper, we study the statistics of MDL in the strong-coupling regime. In this regime, a MMF can be modeled as a cascade of random matrices [7, 27]. By using the central limit theorem for the product of random matrices, we show that MDL follows a limiting distribution that depends only on the number of modes and a single additional parameter. Using known properties of $2 \times 2$ and very large matrices, we state two propositions, which are verified through extensive numerical simulation. These two important results are the following:

**Proposition I**: In the strong-coupling regime, when the overall MDL is small, the distribution of the overall MDL (measured in units of the logarithm of power gain or decibels) is identical to the eigenvalue distribution of zero-trace Gaussian unitary ensemble.

**Proposition II:** In the strong-coupling regime, when the overall MDL is small, the standard deviation (STD) of the overall MDL $\sigma_{mdl}$ depends solely on the square-root of the accumulated MDL variance $\xi$ via:

$$\sigma_{mdl} = \xi\sqrt{1+\tfrac{1}{12}\xi^2} \ . \tag{1}$$

If the MMF comprises $K$ independent, statistically identical sections, each with MDL variance $\sigma_g^2$, we have $\xi = \sqrt{K}\sigma_g$. In Eq. (1), $\sigma_{mdl}$ and $\xi$ are measured in units of the logarithm of power gain.[1]

Proposition I describes the shape of the MDL distribution, whose spread depends on the variance of the Gaussian unitary ensemble. Proposition II quantifies the spread of the distribution by specifying its STD. Using these propositions, the distribution of MDL can be completely specified by the number of modes and the square-root of the accumulated MDL variance $\xi = \sqrt{K}\sigma_g$.

Propositions I and II have not been proven rigorously. Using numerical simulation, we demonstrate that in the small-MDL region, the shape of the overall MDL distribution essentially depends only on the dimension of the zero-trace Gaussian unitary ensemble, which is the number of modes. For systems with overall MDL in the range of practical interest, $\sigma_{mdl} \leq 12$ dB or $\xi = \sqrt{K}\sigma_g \leq 10$ dB, the overall MDL (measured in decibels or log power gain) has a distribution very close to the eigenvalue distribution of zero-trace Gaussian unitary ensemble.

In Proposition II, the STD of the overall MDL depends solely on $\xi = \sqrt{K}\sigma_g$. For modal dispersion in MMF, or for polarization-mode dispersion (PMD) in SMF [28-30], the overall modal group delay spread depends *linearly* on $\sqrt{K}\sigma_\tau$, where $\sigma_\tau$ is the STD of the group delay spread per section [7, 28-31]. For MDL in MMF, the overall MDL depends *nonlinearly* on $\xi = \sqrt{K}\sigma_g$, as described by Eq. (1). The nonlinear relationship of Proposition II cannot be proven rigorously in the general case, but related results can be derived analytically in the limit of many modes. Proposition II has been tested by comparing Eq. (1) to numerical simulations. The approximation Eq. (1) is highly accurate for two-mode fibers with $\xi = \sqrt{K}\sigma_g$ up to 10 dB, and the region of its validity increases with an increasing number of modes.

A statistical characterization of MDL can be used to compute information-theoretic channel capacities of MMF with MDL. Instead of using brute-force simulation of a cascade of many independent fiber sections described by random matrices, a zero-trace Gaussian unitary ensemble can be directly generated numerically, and its eigenvalues computed. Alternatively, the known joint probability density of a zero-trace Gaussian unitary ensemble, as given in [7], can be employed. We demonstrate that computations of both average capacities and outage capacities by these methods match those by brute force simulation in the region of small MDL.

---

[1] Quantities measured in units of log power gain can be converted to decibels by multiplying by $\gamma = 10/\ln10 \approx 4.34$, e.g., $\sigma_{mdl}$ (dB) = $\gamma\sigma_{mdl}$ (log power gain). Unless noted otherwise, all expressions in this paper assume that both $\xi$ and $\sigma_{mdl}$ are expressed in log power gain units.

The remainder of this paper is organized as follows. Sec. 2 presents the matrix model of MMF with strong mode coupling and provides arguments for Propositions I and II. Sec. 3 presents numerical simulations verifying Propositions I and II. Sec. 4 computes channel capacities of MMF with MDL and compares them to results computed for the zero-trace Gaussian unitary ensemble. Secs. 5 and 6 are discussion and conclusions, respectively.

## 2. Matrix Model for Multimode Fiber with Strong Mode Coupling

The model presented here is valid in the strong-coupling regime, where the overall fiber length is much greater than the correlation length over which the MDL profile can be considered constant, such that the fiber can be modeled as a concatenation of many independent sections. The model presented here is valid regardless of the actual correlation length or the statistics of MDL in each individual section.

### 2.1 Random Matrix Model

In the strong-coupling regime, a MMF is divided into $K$ sections, with each section modeled as a random matrix. The length of each section should be at least the correlation length, such that each section can be considered independent of the others [7, 27].

Assume that the MMF supports $D$ propagating modes.[2] At an angular frequency $\omega$, the transfer characteristic of the $k$th section may be modeled by a matrix $\mathbf{M}^{(k)}(\omega)$, which is the product of three $D \times D$ matrices:

$$\mathbf{M}^{(k)}(\omega) = \mathbf{V}^{(k)} \mathbf{\Lambda}^{(k)}(\omega) \mathbf{U}^{(k)*}, \quad k = 1, \ldots, K, \quad (2)$$

where $*$ denotes Hermitian transpose, $\mathbf{U}^{(k)}$ and $\mathbf{V}^{(k)}$ are frequency-independent random unitary matrices representing modal coupling at the input and output, respectively, and $\mathbf{\Lambda}^{(k)}(\omega)$ is a diagonal matrix representing modal propagation in the $k$th section. Including both MDL and modal dispersion, $\mathbf{\Lambda}^{(k)}(\omega)$ can be expressed as:

$$\mathbf{\Lambda}^{(k)}(\omega) = \mathrm{diag}\left[ e^{\frac{1}{2}g_1^{(k)} - j\omega\tau_1^{(k)}}, e^{\frac{1}{2}g_2^{(k)} - j\omega\tau_2^{(k)}}, \ldots, e^{\frac{1}{2}g_D^{(k)} - j\omega\tau_D^{(k)}} \right], \quad (3)$$

where, in the $k$th section, the vector

$$\mathbf{g}^{(k)} = \left( g_1^{(k)}, g_2^{(k)}, \ldots, g_D^{(k)} \right) \quad (4)$$

describes the uncoupled MDL, and $\boldsymbol{\tau}^{(k)} = \left( \tau_1^{(k)}, \tau_2^{(k)}, \ldots, \tau_D^{(k)} \right)$ describes the uncoupled modal groups delays. Neither the mean MDL, nor the mean group delay, affect the statistical properties of interest, so without loss of generality, we assume that $\tau_1^{(k)} + \tau_2^{(k)} + \cdots + \tau_D^{(k)} = 0$ and $g_1^{(k)} + g_2^{(k)} + \cdots + g_D^{(k)} = 0$.

The overall transfer matrix of a MMF having $K$ sections is:

$$\mathbf{M}^{(t)} = \mathbf{M}^{(K)} \cdots \mathbf{M}^{(2)} \mathbf{M}^{(1)} . \quad (5)$$

At each frequency, the overall MDL is described by the singular values of $\mathbf{M}^{(t)}$ or, equivalently, by the eigenvalues of $\mathbf{M}^{(t)}\mathbf{M}^{(t)*}$ or $\mathbf{M}^{(t)*}\mathbf{M}^{(t)}$, which are both Hermitian matrices. Mathematically, the eigenvalues of $\mathbf{M}^{(t)}\mathbf{M}^{(t)*}$ are the squares of the singular values of $\mathbf{M}^{(t)}$. The singular values of $\mathbf{M}^{(t)}$ describe electric field gains, while the eigenvalues of $\mathbf{M}^{(t)}\mathbf{M}^{(t)*}$ describe optical power gains.

---

[2]Throughout this paper, "modes" include both polarization and spatial degrees of freedom. For example, the two-mode case can describe the two polarization modes in SMF.

Similar to multi-input multi-output (MIMO) wireless systems [32, 33], at any single frequency, using singular value decomposition, the overall matrix $\mathbf{M}^{(t)}$ can be decomposed into $D$ spatial channels:

$$\mathbf{M}^{(t)} = \mathbf{V}^{(t)}\mathbf{\Lambda}^{(t)}\mathbf{U}^{(t)*}, \tag{6}$$

where $\mathbf{U}^{(t)}$ and $\mathbf{V}^{(t)}$ are the input and output unitary beam-forming matrices, and we have defined

$$\mathbf{\Lambda}^{(t)} = \text{diag}\left[e^{\frac{1}{2}g_1^{(t)}}, e^{\frac{1}{2}g_2^{(t)}}, \ldots, e^{\frac{1}{2}g_D^{(t)}}\right].$$

Here, $\mathbf{g}^{(t)} = \left(g_1^{(t)}, g_2^{(t)}, \ldots, g_D^{(t)}\right)$ is a vector of the logarithms of the eigenvalues of $\mathbf{M}^{(t)}\mathbf{M}^{(t)*}$, which quantifies the overall MDL of the MIMO system. Our goal here is to study the statistics of the overall MDL described by $\mathbf{g}^{(t)}$.

In an individual section, say the $k$th section, the uncoupled MDL $\mathbf{g}^{(k)}$ and group delay $\mathbf{\tau}^{(k)}$ are modeled as independent of frequency within the band of a signal of interest. In Eq. (5), multiplication of many matrices of the form Eq. (2) causes the overall MDL and overall group delay to become frequency-dependent. In the decomposition given by Eq. (6), $\mathbf{\Lambda}^{(t)}$, $\mathbf{U}^{(t)}$, and $\mathbf{V}^{(t)}$ are all frequency-dependent. We study the overall MDL at a single frequency, and suppress the frequency dependence in the remainder of this paper.

In operation of a MIMO system, as represented by Eq. (5), the receiver estimates the overall channel matrix $\mathbf{M}^{(t)}$, computes the channel decomposition using Eq. (6), sends a description of the beam-forming matrix $\mathbf{U}^{(t)}$ and the gain vector $\mathbf{g}^{(t)}$ to the transmitter, and uses the receive beam-forming vector $\mathbf{V}^{(t)}$ in decoding received signals [33]. The matrix $\mathbf{U}^{(t)}$ and the gain vector $\mathbf{g}^{(t)}$ represent channel state information (CSI) that is to be fed back from the receiver to the transmitter. In the bit-loading process, both the power and information bits in each spatial channel may be allocated according to the gain vector $\mathbf{g}^{(t)}$. In the general case that the channel matrix $\mathbf{M}^{(t)}$ is frequency-dependent, frequency-dependent channel decomposition [given by Eq. (6)] and bit loading may be performed. When using OFDM, the bit loading may be performed for each subcarrier or for a group of adjacent subcarriers. Using single-carrier modulation, frequency-dependent channel decomposition and bit loading is possible, but becomes complex to implement in the regime of strong mode coupling [34].

In a long-haul system, the feedback process described above may become impractical if the MMF changes on a time scale shorter than or comparable to the round-trip propagation delay. When feedback becomes impossible, space-time codes [35-36] or error-correction codes across the spatial channels can provide diversity. In these cases, all spatial channels are allocated equal powers.

*2.2 Measures of Mode-Dependent Loss and Gain*

We are now prepared to define properly two terms used in this paper to quantify MDL: accumulated MDL and overall MDL.

Accumulated MDL refers to the sum of the uncoupled MDL values in all $K$ sections comprising a fiber. The variance of the accumulated MDL is:

$$\xi^2 = \sigma_{g^{(1)}}^2 + \sigma_{g^{(2)}}^2 + \cdots + \sigma_{g^{(K)}}^2, \tag{7}$$

where $\sigma_{g^{(k)}}^2$, $k = 1, \cdots, K$, are the variances of the uncoupled MDL vectors in the individual sections. If the individual sections are statistically identical, we have $\xi = \sqrt{K}\sigma_g$, where $\sigma_g^2$ is the variance of the uncoupled MDL in each section.

Overall MDL refers to the end-to-end coupled MDL of a fiber comprising $K$ sections. The overall MDL is computed from the gain vector $\mathbf{g}^{(t)}$ that appears in Eq. (6). The gains in $\mathbf{g}^{(t)}$ are assumed to be ordered as $g_1^{(t)} \geq g_2^{(t)} \geq \cdots \geq g_D^{(t)}$, and are assumed to sum to zero: $g_1^{(t)} + g_2^{(t)} + \cdots + g_D^{(t)} = 0$. When taken together, the $D$ elements of $\mathbf{g}^{(t)}$ have zero mean; equivalently, an element $g_i^{(t)}$ chosen randomly from $\mathbf{g}^{(t)}$ has zero mean.

Two statistical parameters are important for characterizing MDL: the STD of overall MDL $\sigma_{\mathrm{mdl}}$ and the mean of the maximum MDL difference $E\{g_1^{(t)} - g_D^{(t)}\}$, where $E\{\ \}$ denotes the expectation of a random variable.

The STD of overall MDL is computed over all $D$ elements of the gain vector $\mathbf{g}^{(t)}$; equivalently, it is the square-root of $\sigma_{\mathrm{mdl}}^2 = E\{(g_i^{(t)})^2\}$, where $g_i^{(t)}$ is an element chosen randomly from $\mathbf{g}^{(t)}$. The STD of overall MDL $\sigma_{\mathrm{mdl}}$ has the same units as the gains in $\mathbf{g}^{(t)}$.

The mean of the maximum MDL difference, $E\{g_1^{(t)} - g_D^{(t)}\}$, quantifies the gain difference between the strongest and weakest modes. In a two-mode fiber, $E\{g_1^{(t)} - g_2^{(t)}\}$ is commonly referred to as the mean polarization-dependent loss (PDL).

The accumulated MDL variance $\xi^2$, given by Eq. (7), does not equal the variance of overall MDL because of the nonlinearity inherent in Proposition II, given by Eq. (1). Much of the complexity of PDL and MDL arises from the nonlinearity of Eq. (1).

*2.3 Properties of the Product of Random Matrices*

Characterizing the statistics of the singular values of $\mathbf{M}^{(t)}$, or those of the eigenvalues of $\mathbf{M}^{(t)}\mathbf{M}^{(t)*}$, is the key to understanding the performance of MDM systems, as in MIMO wireless systems [32-33]. The analysis here is complicated by the fact that $\mathbf{M}^{(t)}$ is the product of random matrices [see Eq. (5)], rather than the sum of random matrices, as in [7]. Since the early works [37] and [38], there have been many studies on the statistics of the products of random matrices, but most addressed the Lyapunov exponent of the products [39, 40]. We are interested here in the statistics of the eigenvalues of a product of matrices, not the Lyapunov exponent.

Because matrix multiplication is not commutative, i.e., $\mathbf{AB}$ is not generally equal to $\mathbf{BA}$, even for square matrices, the logarithm of the product of two matrices, $\log \mathbf{AB}$, is not equal to the sum of $\log \mathbf{A}$ and $\log \mathbf{B}$. Unlike the product of positive random variables (which do commute), which has its central limit as the log-normal distribution, the product of positive random matrices (those with positive eigenvalues) does not generally have its central limit as the exponent of a Gaussian unitary ensemble. The central limit theorem for the summation of random matrices is not necessary helpful for the understanding of the products of random matrices.

For any matrix $\mathbf{X}$ and a very small number $\delta$, we have $\log(\mathbf{I} + \delta \mathbf{X}) \approx \delta \mathbf{X}$, where $\mathbf{I} + \delta \mathbf{X}$ is intended to describe a matrix $\mathbf{M}^{(k)}$ when the gain vector $\mathbf{g}^{(k)}$ has small norm. If both matrices $\mathbf{A}$ and $\mathbf{B}$ are positive and both $\log \mathbf{A}$ and $\log \mathbf{B}$ are small, $\log \mathbf{AB} \approx \log \mathbf{A} + \log \mathbf{B}$. As an approximation, the product of positive random matrices with small logarithm has a central limit as the exponential of a Gaussian ensemble. When applied to the overall product matrix $\mathbf{M}^{(t)}$, if all gain vectors $\mathbf{g}^{(k)}$ are small, the matrix $\mathbf{M}^{(t)}$ is the exponential of the Gaussian ensemble. The approximation used here is similar to the results of Berger [41]. This small-gain approximation yields Proposition I, but not Proposition II. If we made the approximation $\log(\mathbf{I} + \delta \mathbf{X}) \approx \delta \mathbf{X}$, Proposition II would be a linear relationship, unlike Eq. (1).

Of course, the approximation Eq. (1) is linear when $\xi$ is far less than unity. Numerical simulation shows that Proposition I remains valid for $\xi$ up to 10 dB (about 2.3 in log power gain units), a regime in which, equivalently speaking, $\log \mathbf{A}$ and $\log \mathbf{B}$ are not very small.

For random matrices of the form Eq. (2), at a single frequency, we are interested in the statistics of the singular values of the product Eq. (5) in the decomposition Eq. (6). Approximate results were obtained for the case that Eq. (2) is $2 \times 2$ in the study of PDL in SMF [42-43], and for the case that Eq. (2) has very large size in the study of free random variables [44-45]. In both cases, for positive matrices $\mathbf{A}$ and $\mathbf{B}$, the product $\mathbf{AB}$ may be interpreted as $\mathbf{A}^{1/2}\mathbf{B}\mathbf{A}^{1/2*}$, similar to free probability theory. The repetition of $\mathbf{A}^{1/2}\mathbf{B}\mathbf{A}^{1/2*}$ with $\mathbf{A} = \mathbf{M}^{(2)}\mathbf{M}^{(2)*}$ and $\mathbf{B} = \mathbf{M}^{(1)}\mathbf{M}^{(1)*}$ from $k = 2$ to $K$ yields $\mathbf{M}^{(t)}\mathbf{M}^{(t)*}$.

PDL in a SMF [42] is modeled using $2 \times 2$ matrices. Both [42] and [46] showed that for low PDL, the PDL (measured in decibels or log power gain units) has a Maxwellian distribution, the same as the distribution of the group delay in SMF with PMD [28-30]. In a recent study [7], we showed that the Maxwellian is the eigenvalue distribution for a zero-trace $2 \times 2$ unitary Gaussian ensemble. The zero-trace $2 \times 2$ unitary Gaussian ensemble may be obtained by the summation of many zero-trace $2 \times 2$ Hermitian matrices [7].

To compare the results of [42] and [46] for the products of $2 \times 2$ random matrices and the results of [7] for the summation of $2 \times 2$ random matrices, in terms of its eigenvalues, the central limit for the product of random matrices has the same distribution shape as the central limit for the summation of random matrices. This statement is only valid in the small-PDL limit, but the Maxwellian distribution has been found to be accurate even for large PDL [43]. The model used here is largely the same as that in [46], except that the $2 \times 2$ matrices $\mathbf{U}^{(k)}$ and $\mathbf{V}^{(k)}$ are complex unitary matrices instead of the real orthogonal matrices in [46].

Free random variables are equivalent to large matrices of the form Eq. (2) [44-45], and the statistics of free random variables are the statistical properties of the eigenvalues of the large matrices. The central limit theorem for the summation of independent free random variables gives a semicircle distribution [47], similar to the distribution of the eigenvalues of a large class of large random matrices [48]. In free probability theory, the semicircle distribution serves a function analogous to the normal distribution, which is the central limit for the summation of random variables in traditional probability theory [49, Sec. 8-5].

In traditional probability theory, the product of independent positive random variables has a central limit as the log-normal distribution. The log-normal distribution models shadowing in wireless systems [50, Sec. 3.2.9] and distortion by stimulated Raman scattering in optical fiber systems [51]. As shown in [52], the log-semicircle distribution is found to be the central limit of the product of positive free random variables if the free random variables have small variance; in the notation used here, this corresponds to $\xi$, defined in Eq. (7), having a small value. Although there have been many studies on the properties of the products of free random variables [53-55], to our knowledge, none have studied the distribution explicitly.

From the results of [52], equivalently speaking, when the MDL is small and in the strong-coupling regime, the overall MDL has a log-semicircle distribution. In [7], it is shown that the modal group delays follow a semicircle distribution in the limit of a large number of modes. MDL (measured in decibels or log power gain) has the same distribution as the modal group delays in the limit of a large number of modes. Expression Eq. (1) is basically equivalent to the results in [52], which were derived from the free probability theory.

Based on the discussion above, Proposition I is correct both for $2 \times 2$ and for large matrices, i.e., for SMF with PDL and for a many-mode fiber with MDL. Proposition II is derived from the theory of free random variables. Free random variables are large random matrices that are used to model MMF with a large number of modes. These arguments do not represent rigorous proofs in the general case for Propositions I and II, however. In the next section, numerical simulation is used to justify extending Propositions I and II to the general case, in which the number of modes $D$ ranges from 2 to infinity.

## 3. Numerical Simulations of Mode-Dependent Loss and Gain

In this section, we use numerical simulation to verify Propositions I and II. We start by addressing PDL in a two-mode fiber, as analytical approximations are well-known. We then address the many-mode case, comparing to analytical results from free-probability theory. Finally, we address fibers with four and eight modes, to further test our analytical approximations.

*3.1 Two-Mode Fiber*

The simplest possible case, a two-mode fiber, describes PDL in SMF, where approximate analytical results were derived for the small-PDL regime [42] and extended to the large-PDL regime [43]. To facilitate comparison of our results with the previous PDL literature, we will quantify MDL in two different ways in the two-mode case. In order to include fibers with more than two modes, we have defined the STD of the overall MDL $\sigma_{\mathrm{mdl}}$ by Eq. (1). In the two-mode case, $\sigma_{\mathrm{mdl}}$ is the STD of $g_1^{(t)}$, $g_2^{(t)}$ taken together; equivalently, it is the root-mean-square (RMS) value of either $g_1^{(t)}$ or $g_2^{(t)}$. We define the mean MDL difference as the mean of $g_1^{(t)} - g_2^{(t)} = 2g_1^{(t)} = -2g_2^{(t)}$, which corresponds to the mean PDL as defined in [42-43].

In [42], instead of using Eq. (1), the STD of overall MDL was approximated by

$$\sigma_{\mathrm{mdl}} = \frac{3}{\sqrt{2}} \sqrt{\exp\left(\tfrac{2}{9}\xi^2\right) - 1}, \tag{8}$$

where both $\sigma_{\mathrm{mdl}}$ and $\xi$ were defined as in Eq. (1). The STD of overall MDL given by Eq. (8) is the same as that in [42]. Power gain is used here instead of the electric field gain used in [42]. The mean MDL difference in [42] is twice the STD of MDL $\sigma_{\mathrm{mdl}}$ as defined here. To draw the correspondence between our notation and that of [42]: $\xi^2 = z$ and $\sigma_{\mathrm{mdl}}^2 = \langle \rho^2 \rangle / \gamma^2$.

From [42], the probability density of PDL (in decibels or log power gain) is the well-known Maxwellian distribution, which is the same as the distribution of differential group delay in SMF with PMD [28-30]. Fig. 1(a) shows the simulated probability density of the MDL of a two-mode fiber compared with the two-sided Maxwellian distribution given in Table I. The probability densities in Fig. 1(a) are shown with the *x*-axis normalized by the simulated STD of the overall MDL $\sigma_{\mathrm{mdl}}$. The fiber has $K = 256$ sections. Each independent fiber section has a gain of $\pm\alpha$, where $\alpha = \xi/\sqrt{K} = \sigma_g$, with values of $\xi$ labeled in Fig. 1(a). The unitary matrices $\mathbf{U}^{(k)}$ and $\mathbf{V}^{(k)}$ are generated by the method described in the Appendix. The eigenvalues of the cascaded matrix $\mathbf{M}^{(t)}\mathbf{M}^{(t)*}$ are found numerically. The probability densities in Fig. 1(a) are each obtained from 200,000 eigenvalues. In Fig. 1(a), the simulated probability density matches the two-sided Maxwellian distribution very well for $\xi \leq 10$ dB.

Fig. 1(b) shows the STD of the overall MDL $\sigma_{\mathrm{mdl}}$ as a function of the square-root of the accumulated MDL variance $\xi = \sqrt{K}\sigma_g$. Also shown in Fig. 1(b) is the mean MDL difference $E\{g_1^{(t)} - g_2^{(t)}\}$, which is approximately twice ($2\sqrt{3/\pi}$ times, to be exact) the STD of the overall MDL $\sigma_{\mathrm{mdl}}$.

In Fig. 1(b), we observe that the approximation of $\sigma_{\mathrm{mdl}}$ given by Eq. (8) [42] is always larger than the simulated results, but is valid within 0.5 and 1 dB up to $\xi = 8$ and 9 dB, respectively. The approximation of $\sigma_{\mathrm{mdl}}$ given by Eq. (1) is always smaller than the simulated results, but is valid within 0.5 and 1 dB up to $\xi = 9$ and 12 dB, respectively. While the approximation of $\sigma_{\mathrm{mdl}}$ given by Eq. (8) [42] deviates from the simulated results beyond $\xi = 10$ dB, that given by Eq. (1) lies within 3 dB up to $\xi = 20$ dB.

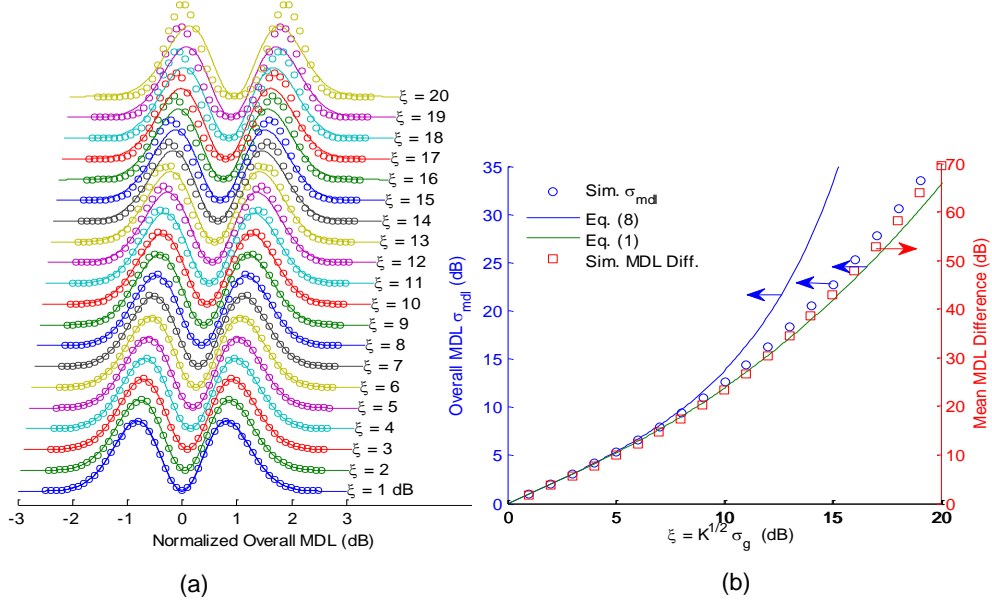

Fig. 1. MDL in a two-mode fiber. (a) Comparing the simulated distribution of the overall MDL with the two-sided Maxwellian distribution given in Table I. The *x*-axis for each curve is normalized by the simulated STD of the overall MDL $\sigma_{mdl}$. (b) Comparing the simulated STD of overall MDL with approximations (1) and (8). The simulated mean MDL difference is also shown for comparison.

As explained in [42], in practical SMF systems, the mean MDL difference (called mean PDL in [42]) should be much less than about 15 dB, corresponding to $\xi = \sqrt{K}\sigma_g$ less than about 7 dB in Fig. 1(b). Our purpose for including large values of MDL in Fig. 1 is to test the validity of Eqs. (1) and (8) in predicting the STD of overall MDL $\sigma_{mdl}$. The more exact model of [43], which is applicable beyond the low-MDL regime, is not shown in Fig. 1(b). In [43], the STD of overall MDL is given as $\sigma_{mdl} = \xi\sqrt{1 + \xi^2/9}$, which is very close the approximation given by Eq. (1).

We conclude, similar to [42], that in two-mode fibers having mean MDL differences less than 15 dB, the MDL (measured in decibels or log power gain) follows a Maxwellian distribution, and the approximation Eq. (1) for the STD of the overall MDL is valid. In two-mode fibers, Proposition I is valid for $\xi$ up to 10 dB, corresponding to the STD of overall MDL $\sigma_{mdl}$ up to 13.8 dB and mean MDL difference up to 23.4 dB. The STD of overall MDL, Eq. (1) from Proposition II, is valid for $\xi$ up to about 15 dB.

### 3.2 Fibers with Large Numbers of Modes

For fibers with a large number of modes, the modal group delays have the same statistical properties as the eigenvalues of a large zero-trace Gaussian unitary ensemble [7]. The eigenvalues of large random matrices have a semicircle distribution, as shown by Wigner [56], and shown to be valid for a large class of large random matrices [48]. The summation of free random variables also gives a semicircle distribution, as known from free probability theory [47].

**Table I: Probability distribution of normalized MDL having unit variance.**

| D | Probability density |
|---|---|
| 2 | $3\sqrt{\dfrac{3}{2\pi}}x^2\exp\left(-\dfrac{3}{2}x^2\right)$ |
| 3 | $\sqrt{\dfrac{2}{\pi}}\exp(-2x^2)\left(6x^4 - 3x^2 + \dfrac{5}{8}\right)$ |
| 4 | $\sqrt{\dfrac{10}{\pi}}\exp\left(-\dfrac{5}{2}x^2\right)\left(\dfrac{500}{81}x^6 - \dfrac{200}{27}x^4 + \dfrac{25}{9}x^2 + \dfrac{5}{54}\right)$ |
| 5 | $\sqrt{\dfrac{3}{\pi}}\exp(-3x^2)\left(\dfrac{3375}{128}x^8 - \dfrac{3375}{64}x^6 + \dfrac{8775}{256}x^4 - \dfrac{1185}{256}x^2 + \dfrac{903}{2048}\right)$ |
| 6 | $\sqrt{\dfrac{14}{\pi}}\exp\left(-\dfrac{7}{2}x^2\right)\left(\dfrac{453789}{15625}x^{10} - \dfrac{259308}{3125}x^8 + \dfrac{256221}{3125}x^6 - \dfrac{17493}{625}x^4 + \dfrac{4557}{1250}x^2 + \dfrac{322}{3125}\right)$ |
| 7 | $\sqrt{\pi}\exp(-4x^2)\left(\dfrac{8605184}{32805}x^{12} - \dfrac{2151296}{2187}x^{10} + \dfrac{326536}{243}x^8 - \dfrac{558404}{729}x^6 + \dfrac{361375}{1944}x^4 - \dfrac{30317}{2592}x^2 + \dfrac{88175}{124416}\right)$ |
| 8 | $\sqrt{\dfrac{2}{\pi}}\exp\left(-\dfrac{9}{2}x^2\right)\left(\dfrac{13060694016}{28824005}x^{14} - \dfrac{8707129344}{4117715}x^{12} + \dfrac{3083774976}{823543}x^{10} - \dfrac{2517744384}{823543}x^8 + \dfrac{140457888}{117649}x^6 - \dfrac{23059728}{117649}x^4 + \dfrac{1533654}{117649}x^2 + \dfrac{502065}{1647086}\right)$ |
| ∞ | $\dfrac{1}{2\pi}\sqrt{4-x^2}$, $|x| \le 2$ and zero otherwise. |

Fig. 2(a) shows the simulated eigenvalue distribution of a fiber with 64 modes. Similar to the simulation in Fig. 1(a), the fiber has $K = 256$ sections and all unitary matrices are generated by the method described in the Appendix. Each curve of Fig. 2(a) is constructed using 64,000 eigenvalues. Each MMF section has a modal gain profile $g_i^{(k)} = \pm\alpha$, where $\alpha = \sigma_g$. The first 32 modes have gains of $+\alpha$, and the last 32 modes have gains of $-\alpha$, such that the gains sum to zero. The x-axis of each curve in Fig. 2(a) is normalized by the simulated STD of the overall MDL $\sigma_{mdl}$.

In Fig. 2(a), the simulated eigenvalue distribution is very close to the semicircle distribution given by Table I up to $\xi = 15$ dB. Comparing Figs. 1(a) and 2(a), the simulated distribution in Fig. 2(a) is closer to the semicircle distribution than the simulated distribution in Fig. 1(a) is to the Maxwellian distribution over a larger range of MDL values.

Fig. 2(b) compares the simulated STD of the overall MDL $\sigma_{mdl}$ as a function of $\xi = \sqrt{K}\sigma_g$ to the approximation for $\sigma_{mdl}$ given by Eq. (1) for a fiber with $D = 64$ modes. The approximation Eq. (1) agrees with simulated results within 0.01 dB for $\xi$ up to 10 dB. For $\xi$ from 10 to 20 dB, the overall MDL approximation Eq. (1) is always smaller than the simulated results and the discrepancy between the simulations and the approximation Eq. (1) increases to 0.15 dB. The discrepancy may arise from our simulating matrices of size $D = 64$ instead of infinitely large matrices. The discrepancy may also be caused by numerical uncertainty. Nevertheless, the approximation Eq. (1) can be considered highly accurate.

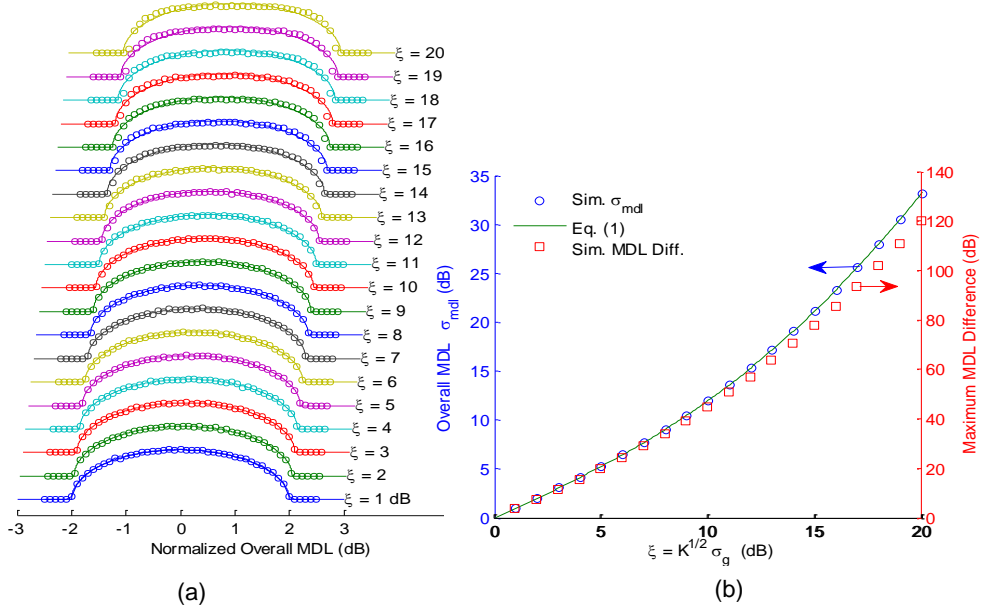

Fig. 2. MDL in MMF with $D = 64$ modes. (a) Comparing the simulated distribution of the overall MDL with the semicircle distribution given in Table I. The *x*-axis for each curve is normalized by the simulated STD of the overall MDL $\sigma_{mdl}$. (b) Comparing the simulated STD of overall MDL with approximation (1). The simulated mean maximum MDL difference is also shown for comparison.

Fig. 2(b) also shows the simulated maximum MDL difference, which is the mean of the maximum gain difference $g_1^{(t)} - g_{64}^{(t)}$. In the range of $\xi$ up to 15 dB, where the simulated distribution is close to the semicircle distribution [as seen in Fig. 2(a)], the STD of overall MDL $\sigma_{mdl}$ is as high as 33.4 dB and the maximum MDL difference is as high as 81 dB. Practical MDM system should have MDL well below those values.

We conclude that for fibers with 64 modes, Proposition I is valid for $\xi$ up to 15 dB, corresponding to $\sigma_{mdl}$ up to 21.2 dB. The overall MDL approximation Eq. (1) from Proposition II is valid for values of $\xi$ up to 20 dB.

*3.3 Other Few-Mode Fibers*

Figs. 1 and 2 confirm numerically that Propositions I and II are valid for fibers with two and 64 modes. Using [7], the eigenvalues of small size zero-trace Gaussian unitary ensembles can be derived analytically by direct integration. Table I shows those distributions with normalized variances of unity.

Figs. 1(a) and 2(a) show that the simulated distribution is close to the theoretical distribution in the region of small $\xi$ and small overall MDL. The approximate STD of overall MDL given by Eq. (1) and the simulation results become closer with an increase in the number of modes. The distributions in Table I are accurate up to certain values of $\xi$, which increase with an increase in the number of modes. The discrepancy between the approximation Eq. (1) and simulated values decreases with an increase in the number of modes.

Figs. 3 and 4 compare the simulated results for $D = 4$ and 8 modes with the theoretical distribution in Table I and the approximation Eq. (1). Each of the curves in Figs. 3(a) and 4(a) is constructed using 200,000 and 400,000 eigenvalues, respectively.

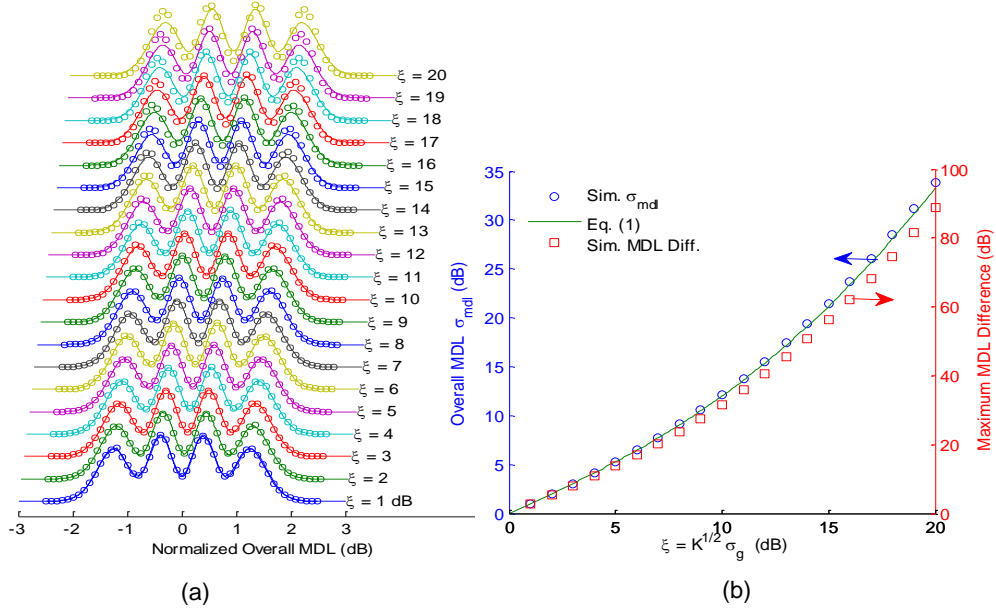

Fig. 3. MDL in MMF with $D = 4$ modes. (a) Comparing the simulated distribution of the overall MDL with the four-peak distribution given in Table I. The $x$-axis for each curve is normalized by the simulated STD of the overall MDL $\sigma_{mdl}$. (b) Comparing the simulated STD of overall MDL with approximation (1). The simulated mean maximum MDL difference is also shown for comparison.

In Fig. 3(a), for a fiber with $D = 4$ modes, we observe that the simulated distribution matches that of Table I until $\xi = 13$ dB, a value slightly higher than for $D = 2$ modes [Fig. 1(a)], but slightly smaller than for $D = 64$ modes [Fig. 2(a)]. As explained in [7], the number of peaks is the same as the number of modes, and each peak corresponds to values where the gain is concentrated. The simulated central two peaks in Fig. 3(a) actually match the theory until $\xi = 17$ dB, but discrepancies appear in the two side peaks.

In Fig. 3(b), for a fiber with $D = 4$ modes, we observe that the approximation for $\sigma_{mdl}$ Eq. (1) is always smaller than the simulated results. The discrepancy is up to 0.11 dB for $\xi$ up to 10 dB but increases to 0.55 dB for $\xi$ up to 20 dB.

In Fig. 4(a), for a fiber with $D = 8$ modes, we observe that the simulated distributions match those in Table I until $\xi = 14$ dB, slightly higher than for $D = 2$ modes [Fig. 1(a)], but slightly smaller than for $D = 64$ modes [Fig. 2(a)]. Similar to Fig. 3(a), the simulated central six peaks of Fig. 4(a) match the theory until $\xi = 17$ dB, but discrepancies appear in the two side peaks.

In Fig. 4(b), for a fiber with $D = 8$ modes, we observe that the approximation of $\sigma_{mdl}$ Eq. (1) agrees with simulations within 0.02 dB for $\xi$ up to 10 dB. For $\xi$ ranging from 10 to 20 dB, the overall MDL approximation Eq. (1) is always smaller than the simulated results, and the discrepancy increases to a maximum of 0.04 dB. We conclude that in fibers with $D = 4$ and 8 modes, in the range of $\xi$ up to 20 dB, corresponding to a STD of overall MDL up to $\sigma_{mdl} = 33$ dB, the overall MDL approximation Eq. (1) can be considered accurate for practical purposes. Proposition I can be considered accurate in the small-MDL region, in which $\xi$ is less than 10 to 15 dB, corresponding to the STD of overall MDL $\sigma_{mdl}$ less than 12 to 21 dB, with the accuracy increasing as the number of modes increases. The approximation for $\sigma_{mdl}$, given by Eq. (1) in Proposition II, is accurate over a range larger than that in which Proposition I is accurate. Specifically, for fibers with more than 8 modes, the approximation

Eq. (1) is accurate in all cases simulated for ξ up to 20 dB and for $\sigma_{mdl}$ up to 33 dB. For fibers with 2 or 4 modes, the overall MDL approximation Eq. (1) is valid for values of ξ up to 15 and 19 dB, respectively, with a maximum error of 0.5 dB.

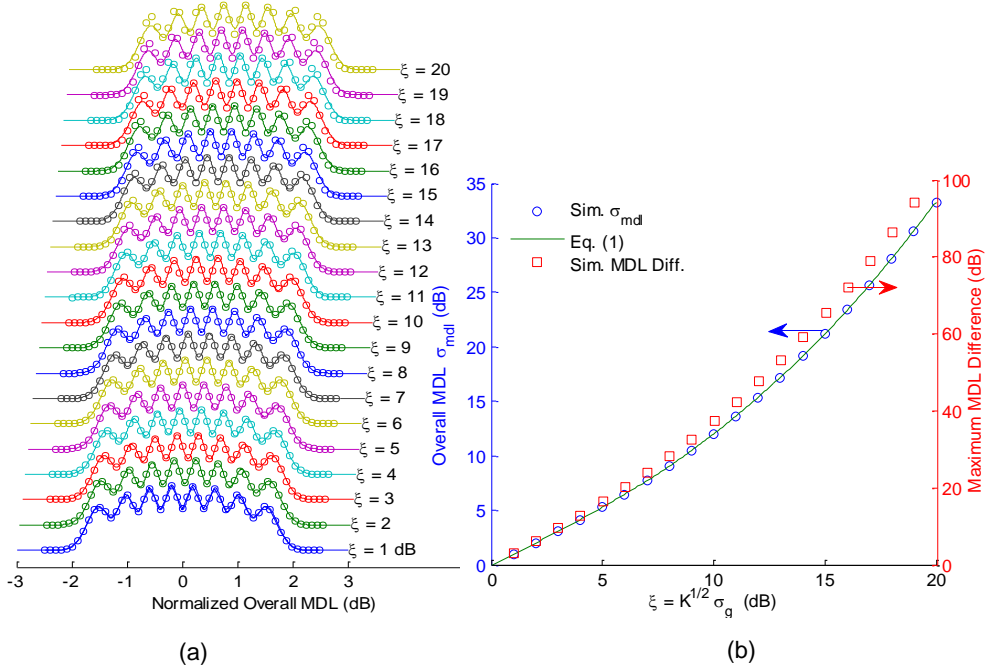

Fig. 4. MDL in MMF with $D = 8$ modes. (a) Comparing the simulated distribution of the overall MDL with the eight-peak distribution given in Table I. (b) Comparing the simulated STD of overall MDL with approximation (1). The simulated mean maximum MDL difference is also shown for comparison.

## 4. Channel Capacity with Mode-Dependent Loss and Gain

An MDM system in MMF is assumed to use coherent detection and inline optical amplification. The dominant noise at the receiver is assumed to arise from amplified spontaneous emission, and the received noise power spectral density is assumed to be the same in each mode, similar to the assumptions in [12-15]. The signal-to-noise ratio (SNR) $\rho_t$ is defined as the received signal power (total over all $D$ modes) divided by the received noise power (per mode).[3] This SNR definition is similar to that in wireless MIMO systems [33].

For a MMF without MDL, all modes have the same gain and the same received power, and the channel capacity is equal to

$$C = D\log_2\left(1+\frac{\rho_t}{D}\right). \qquad (9)$$

Fig. 5 shows the channel capacity as a function of SNR $\rho_t$ for fibers with different numbers of modes. When the SNR is much greater than the number of modes, the capacity increases almost linearly with the number of modes. When the SNR is much less than the number of modes, the capacity is almost independent of the number of modes. In the limit of an infinite number of modes, the capacity is given asymptotically by $C_\infty = \rho_t \log_2 e$, which is independent of the number of modes. Fig. 5 shows clearly that multiple modes increase capacity, especially in systems with sufficiently high SNR.

---

[3] This definition of SNR is compatible with the conservative assumption that the total signal power in all $D$ modes is subject to a constraint independent of $D$, e.g., by fiber nonlinearity or component costs, while the noise power per mode is independent of $D$. Nevertheless, our results, if interpreted correctly, do not depend on this assumption in any way.

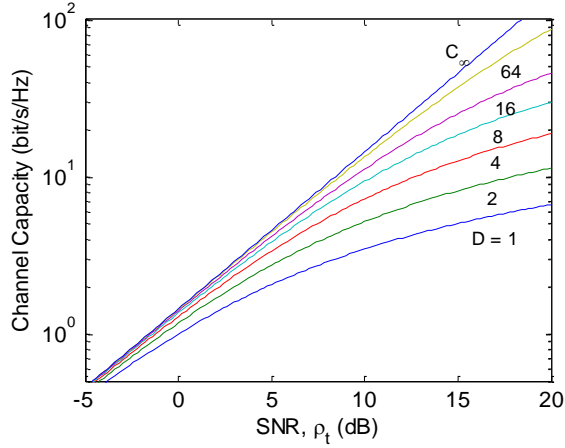

Fig. 5. Channel capacity of a MMF without MDL as a function of SNR $\rho_t$ for various numbers of modes $D$.

The availability of CSI at the transmitter is an essential factor governing channel capacity for MMF with MDL. In the absence of CSI, an increase in MDL always leads to a decrease of capacity. In the extreme limit of MDL, the fiber supports propagation in only one mode. If CSI is available to the transmitter, only the surviving mode is used for transmission, and the channel capacity is $\log_2(1+\rho_t)$, which is Eq. (9) with $D = 1$. If CSI is not available and the surviving mode is not known to the transmitter, the transmitter must allocate equal power to all modes, and the channel capacity is $\log_2(1+\rho_t/D)$. As demonstrated below, channel capacity is improved greatly by the availability of CSI. In some situations, when CSI is available, the channel capacity with MDL may exceed Eq. (9).

*4.1 Average Channel Capacity without Channel State Information*

When CSI is not available, the transmitter allocates equal power to each mode. Given the number of modes $D$ and the STD of overall MDL $\sigma_{mdl}$, the average channel capacity is:

$$C = D \int_{-\infty}^{+\infty} \log_2\left[1+\frac{\chi}{D}\exp(\sigma_{mdl}x)\right]p_D(x)dx , \qquad (10)$$

where $p_D(x)$ is the probability density with unit variance given in Table I, and the constant $\chi$ is determined by the constraint:

$$\chi = \frac{\rho_t}{\int_{-\infty}^{+\infty}\exp(\sigma_{mdl}x)p_D(x)dx} . \qquad (11)$$

If the noise power per mode is normalized to unity, the constant $\chi$ is the total transmitted power and $\chi/D$ in Eq. (10) is the transmitted power per mode.

From Sec. 3, the MDL (measured in units of decibels or log power gain) has zero mean. Measured on a linear scale, the overall power loss/gain of a $K$-section MMF fiber, summed over all $D$ modes, is equal to $\cosh^K \sigma_g$ (as in the simulations shown in Figs. 1 to 4). This overall gain gives a Lyapunov exponent, defined in [38-40], of $\log \cosh \sigma_g$. The factor $\chi$, given by Eq. (11), normalizes the overall gain to unity, as measured on a linear scale. For a system with MDL, the channel gains are not constant, but are random variables. The SNR $\rho_t$ is the mean (statistical average) SNR, and the factor $\chi$ given by (11) can be interpreted as the ratio of the mean SNR to the mean gain of the channel. The product $\chi \exp(\sigma_{\text{mdl}} x)$ in Eq. (10) can be interpreted as the SNR of a channel realization with normalized gain $x$.

Analytical expressions are available for the constant $\chi$, given by Eq. (11), for fibers with the numbers of modes given in Table I, but are too complicated for practical calculations. There is no known analytical expression for Eq. (10) except for the limit of very large $D$. Numerical integration of Eq. (10) may be employed to find the channel capacity. For fibers with a large number of modes, MDL is log-semicircle distributed, and the constant $\chi$ becomes

$$\chi = \frac{\rho_t \sigma_{\text{mdl}}}{2 I_1(\sigma_{\text{mdl}})},$$

where $I_1(\cdot)$ is the modified Bessel function of the first kind.

If $D$ is far larger than $\chi \cdot \exp(2\sigma_{\text{mdl}})$, where $2\sigma_{\text{mdl}}$ is the upper limit of the semicircle distribution, the channel capacity approaches $C_\infty$. In this limit, each mode is allocated a very small power, and the overall capacity is proportional to the total power.

Fig. 6 shows the average capacity with MDL but without CSI, as a function of the square-root of accumulated MDL variance $\xi = \sqrt{K}\sigma_g$, for fibers with various numbers of modes $D$. As CSI is unavailable, equal power is allocated to each mode. The mean SNR is $\rho_t = 10$ dB. In Fig. 6, the theoretical channel capacity is calculated by using Eq. (1) to find the STD of overall MDL $\sigma_{\text{mdl}}$, and the probability distributions in Table I are used in both Eq. (10) and Eq. (11) to compute the average channel capacity. The simulations use channel matrices generated by the same methods used in Figs. 1-4.

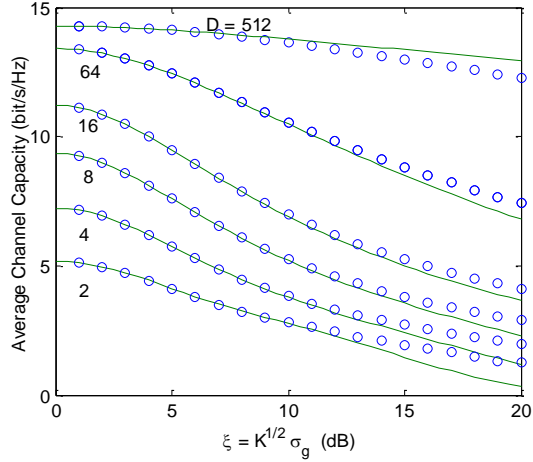

Fig. 6. Average channel capacity of a MMF with MDL and without CSI as a function of $\xi$ for fibers with various numbers of modes $D$. Equal power is allocated to each mode, and SNR is $\rho_t = 10$ dB. Theoretical results are shown as curves and simulated results are shown as circles.

In Fig. 6, in the absence of CSI, average channel capacity always degrades with increasing $\xi$, particularly in fibers with smaller numbers of modes $D$. For values of $\xi$ up to 10 dB, the average channel capacities from theory and simulation match very well. For a two-mode fiber, theoretical and simulated capacities match within 5% up to $\xi = 11$ dB. For a 512-

mode fiber, theoretical and simulated capacities match within 5% up to $\xi$ = 19 dB. The discrepancy between theory and simulation decreases with an increase in the number of modes.

*4.2 Average Channel Capacity with Channel State Information*

When CSI is available at the transmitter, transmit power may be allocated to each mode in an optimal way. If the noise power per mode is normalized to unity and the overall gain vector $\mathbf{g}^{(t)}$ is known, the optimal transmit powers are given by $\left[\mu - e^{-g_i^{(t)}}\right]^+$, $i = 1,...,D$, where $[\ ]^+$ denotes limiting to nonnegative values, i.e., $[x]^+ = \max(0, x)$. The constant $\mu$ is chosen to satisfy the total power constraint:

$$\sum_{i=1}^{D}\left[\mu - e^{-g_i^{(t)}}\right]^+ = \chi, \tag{12}$$

with $\chi$ given by Eq. (11). The average channel capacity is

$$C = E\left\{\sum_{i=1}^{D}\log_2\left(1 + \left[\mu e^{g_i^{(t)}} - 1\right]^+\right)\right\}, \tag{13}$$

which is an expectation over random realizations of the MDL.

A brute-force method to compute Eq. (13) involves generating many random realizations of the overall matrix Eq. (5) and computing their singular values to evaluate the average capacity Eq. (13). Each realization of Eq. (5) is the product of 3$K$ matrices.[4]

Assuming values of MDL sufficiently small to be of practical interest, Propositions I and II are valid (see Sec. 3). In this regime, a more efficient method of computing Eq. (13) is based on Proposition I, namely, that the joint probability density for the vector $\mathbf{g}^{(t)}$ is given by the eigenvalue distribution of a zero-trace Gaussian unitary ensemble [7]. Hence, a zero-trace random Gaussian Hermitian matrix may be generated using the method described in the Appendix, and the eigenvalues of the random matrix can be used to compute Eq. (13). Also, instead of generating new matrices for different values of $\xi$, Proposition II can be exploited. A single matrix may be generated, and the eigenvalues can be scaled using the approximation for $\sigma_{mdl}$, given by Eq. (1).

Fig. 7 shows the average capacity for a system with MDL and with CSI, as a function of the square-root of accumulated MDL variance $\xi = \sqrt{K}\sigma_g$, for fibers with various numbers of modes $D$. The SNR is $\rho_t$ = 10 dB, representing a statistical average. The theoretical curves are obtained using 100,000 zero-trace random Gaussian Hermitian matrices generated as described in the Appendix. Simulated results are obtained using brute-force generation of channel matrices by the same methods used for Figs. 1-4.

In Fig. 7, we see that for fibers with $D$ = 2, 4 or 8 modes, even with CSI, the channel capacity decreases with increasing MDL. With $D$ = 16 modes, at small MDL, the capacity increases with increasing MDL, although it eventually decreases with a further increase in MDL. More generally, MDL can increase capacity when the SNR is small relative to the number of modes $D$. This may seem counter-intuitive, but can be made plausible by the following example for $D$ = 2 modes. Assume that the total transmitted power is unity, the noise power per mode is unity and the mean SNR is $\rho_t$ = 1. Hence, the mean channel gain is 0 dB, corresponding to unity in linear units. Without MDL, using Eq. (9), the channel capacity is 2log$_2$(1+0.5) = 1.17 bit/s/Hz. Now suppose that MDL is present, and that in a particular channel realization, the modal gains are −3.01 dB and +1.76 dB, which correspond to 0.5 and 1.5 in linear units (the mean of these two values is unity, as in the absence of MDL). When CSI is not available, the transmitter allocates powers of 0.5 to each mode. The capacity is $\log_2(1+0.5\cdot 0.5) + \log_2(1+1.5\cdot 0.5) = 1.13$ bit/s/Hz, slightly smaller than without MDL. When CSI is available, the transmitter allocates all the power to the stronger mode. The capacity becomes $\log_2(1+1.5\cdot 1) = 1.32$ bit/s/Hz, slightly larger than that without MDL.

---

[4] The total number of matrices can be reduced to 2$K$ + 1 by combining $\mathbf{U}^{(k+1)}$ and $\mathbf{V}^{(k)}$, $k$ = 1,..., $K$ − 1, into single random unitary matrices.

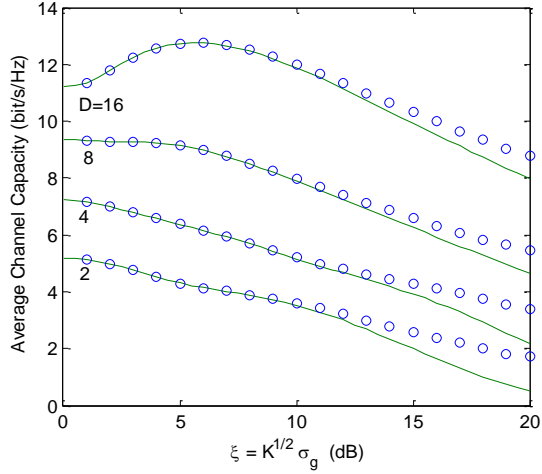

Fig. 7. Average channel capacity of a MMF with MDL and with CSI as a function of ξ for fibers with various numbers of modes *D*. Transmit power is allocated optimally to each mode, and SNR is $\rho_t = 10$ dB. Theoretical results are shown as curves and simulated results are shown as circles.

In Fig. 7, for a two-mode fiber, theoretical and simulated average capacities agree within 5% up to ξ = 11 dB. For a 16-mode fiber, theoretical and simulated average capacities agree within 5% up to ξ = 17 dB. The discrepancy decreases with an increasing number of modes.

*4.3 Outage Capacity*

Outage capacity is often considered a better performance measure than average capacity for channels with random gains, such as in an MDM system with MDL [27]. The outage capacity may be computed using randomly generated Gaussian Hermitian matrices, similar to the method used to compute average capacity with CSI, as described in Sec. 4.2. Such a method is much more efficient than the brute-force multiplication of 3*K* matrices, as given by Eq. (5). With CSI, the outage capacity is computed from a gain vector $\mathbf{g}^{(t)}$ similar to Eq. (13), and is given by:

$$\Pr\left\{\sum_{i=1}^{D}\log_2\left(1+\left[\mu e^{g_i^{(t)}}-1\right]^+\right) \leq C_{\text{out}}\right\} = p_{\text{out}}, \qquad (14)$$

where Pr{ } denotes probability, $C_{\text{out}}$ and $p_{\text{out}}$ are the outage capacity and outage probability, respectively. The constant μ in Eq. (14) is determined by the total power constraint Eq. (12). Without the CSI, the outage probability is given by

$$\Pr\left\{\sum_{i=1}^{D}\log_2\left(1+\frac{\chi}{D}e^{g_i^{(t)}}\right) \leq C_{\text{out}}\right\} = p_{\text{out}},$$

where the constant χ is given by Eq. (11).

Fig. 8 shows the outage capacity of a MMF with MDL, as a function of the square-root of accumulated MDL variance $\xi = \sqrt{K}\sigma_g$, for fibers with various numbers of modes $D$. The outage probability is $p_{\text{out}} = 10^{-3}$, and the SNR is $\rho_t = 10$ dB. Systems with or without CSI are considered. The theoretical curves are obtained using 100,000 zero-trace random Gaussian Hermitian matrices generated, as described in the Appendix. Simulated results are obtained using brute-force generation of channel matrices, as used for Figs. 1-4.

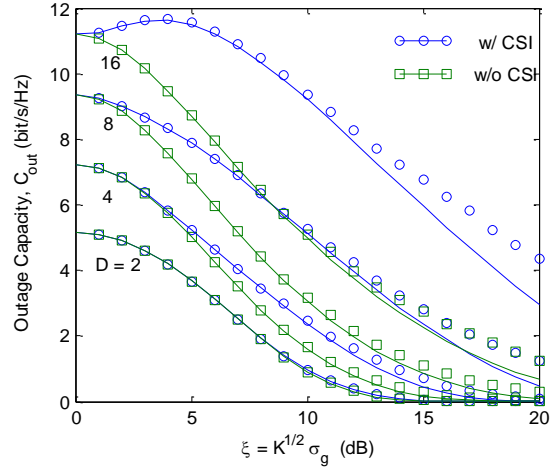

Fig. 8. Outage capacity of a MMF with MDL as a function of $\xi$ for fibers with various numbers of modes $D$. The outage probability is $10^{-3}$, and SNR is $\rho_t = 10$ dB. Theoretical results are shown as curves and simulated results are shown as open symbols. Blue curves/blue circles: with CSI; green curves/green squares: without CSI.

In Fig. 8, we observe that the outage capacity decreases rapidly with increasing MDL for all numbers of modes, with or without CSI.[5] This rapid decrease is caused by a large fraction (or even all) of modes having small gains, and is a consequence of the difference between the average gains measured on linear and decibel scales. The average of the linear gains always exceeds the average of the decibel gains because the arithmetic mean is always larger than the geometric mean.

The rapid decrease of outage capacity with increasing MDL can be explained most easily in the two-mode case, although a similar explanation can be applied to any number of modes. Referring to Fig. 1(a), the origin of the $x$-axis is the mean decibel-scale gain of 0 dB (or the geometric mean linear-scale gain of unity), which is a constant, independent of the value of $\sigma_{\text{mdl}}$. Recall that the stronger and weaker modes always have gains larger and smaller than the origin of Fig. 1(a), respectively. The arithmetic mean of the linear-scale gain is always on the positive side of the origin, and increases with increasing MDL. Now consider a specific numerical example, using only linear gain units. Suppose the MDL is high, and the arithmetic mean gain is 10. Because the geometric mean gain is always unity, the stronger mode has a gain in the range $(1, +\infty)$ and the weaker mode has a gain in the range $(0, 1)$. Suppose that in a particular realization, the stronger and weaker modes have gains of 10 and 0.1, respectively, an average gain close to the arithmetic mean. Assuming the noise level per mode is unity and CSI is not available, the capacity is $\log_2(1+10/2) + \log_2(1+0.1/2) = 2.66$ bit/s/Hz. This is far above the outage capacity. Now suppose that in another realization, the stronger and weaker modes both have gains approaching unity, an average gain close to the geometric mean. Again assuming a noise level of unity and no CSI, the capacity approaches 2log$_2$(1+1/2) = 1.17 bit/s/Hz. This value is very close to the outage capacity.

---

[5] Similar to Fig. 7, with $D = 16$ modes, when CSI is available, the capacity first increases slightly with increasing MDL, then decreases with a further increase in MDL.

In Fig. 8, theoretical and simulated outage capacities match well up to $\xi$ = 10 dB. A difference up to 5% is observed at higher values of $\xi$ as the number of modes increases. In all cases, the theoretical outage capacity lies below simulated values, so the theoretically derived estimates are conservative, particularly at larger values of $\xi$.

## 5. Discussion

Based on [43, 52], the STD of overall MDL is known to be given exactly by

$$\sigma_{mdl} = \xi\sqrt{1 + \frac{1}{\kappa}\xi^2} , \qquad (15)$$

with $\kappa = 9$ for $D$ = 2 and $\kappa = 12$ for $D = \infty$. Using curve fitting, the best-fit values are $\kappa = 11.4$ and 12.0 for $D$ = 4 and 8, as shown in Figs. 3 and 4, respectively. The best-fit value of $\kappa$ in Eq. (15) approaches 12 rapidly with an increasing number of modes, providing a rationale behind our choice of $\kappa = 12$ in Eq. (1).

Optical amplifiers may be subject to saturation, which can cause variations in the signal power and noise level along the propagation path. Any power variations that are uniform across modes will not affect the statistics of MDL. Nevertheless, saturation effects may impact the channel capacities discussed in Sec. 4. The capacities computed in Sec. 4 assume that the total launched power levels, equivalent to $\chi$ given by (11), do not change with the random realizations in MDL. This assumption of constant launched power requires that there be no amplifier saturation caused by MDL. For example, consider a two-mode case with unit total input power, unit noise level (in the absence of saturation), unit SNR, and gains of 0.5 and 1.5 (measured in linear units). In the absence of saturation, the output powers may vary from 0.5 to 1.5, depending on the alignment of the input signal to the principal modes of MDL (i.e., the modes of minimum and maximum gain). When saturation occurs, however, the maximum output powers may be limited to values less than the nominal value of 1.5, while noise levels decrease proportionally to maintain the mean SNR of unity. The model given here does not take account of these effects.

For an input signal vector $\mathbf{x}_0$ and the overall transfer matrix of a $K$-section fiber given by Eq. (5), the theory in [39-40] can be used to characterize the norm $\|\mathbf{M}^{(t)}\mathbf{x}_0\|$, which is log-normal distributed. Considering the case without CSI and assuming the input vector is largely randomly oriented with respect to the principal modes of MDL, after propagating through a few fiber sections, the modal power distribution becomes log-normal. Even when CSI is available, the end-to-end principal modes, given by $\mathbf{U}^{(t)}$, are not the same as the principal modes in the first fiber section, given by $\mathbf{U}^{(1)}$. The overall matrix $\mathbf{U}^{(t)}$ should be correlated with $\mathbf{U}^{(k)}, \mathbf{V}^{(k)}, k = 1,...,K$, but this correlation is not well-known except for very special cases.

In Sec. 4, the noises for the different principal modes are assumed to be independent and identically distributed (i.i.d.). In the strong-coupling regime and with many amplifiers serving as independent sources of amplified spontaneous emission, this assumption should be substantially correct. This can be justified as follows.

For simplicity, suppose the $K$th (last) fiber section contains a noise source, and that in the local eigenmodes of the $K$th section, the noise powers are $\sigma_{K,i}^2$, $i$ = 1, ..., $D$. At the fiber output, the electric fields from this noise source are described by a vector $\mathbf{V}^{(K)}\text{diag}[\sigma_{K,1}, \sigma_{K,2},...,\sigma_{K,D}]\mathbf{n}$, where $\mathbf{n}$ is a $D$-dimensional Gaussian noise vector with unit variance in each element. At the fiber output, the noise correlation matrix is:

$$\mathbf{V}^{(K)}\text{diag}[\sigma_{K,1}^2, \sigma_{K,2}^2,...,\sigma_{K,D}^2]\mathbf{V}^{(K)*} . \qquad (16)$$

For a given realization of the random unitary matrix $\mathbf{V}^{(K)}$, the output noises may be neither independent nor identically distributed. Taking an expectation over all random matrices $\mathbf{V}^{(K)}$ yields:

$$E\left\{\mathbf{V}^{(K)}\text{diag}\left[\sigma_{K,1}^2, \sigma_{K,2}^2, \ldots, \sigma_{K,D}^2\right]\mathbf{V}^{(K)*}\right\} = \frac{\sigma_{K,1}^2 + \sigma_{K,2}^2 + \cdots + \sigma_{K,D}^2}{D}\mathbf{I}, \quad (17)$$

which is a constant times the identity matrix, and hence describes a noise vector with i.i.d. elements.

Considering a noise source in the *k*th fiber section, its output noise contribution is
$$\mathbf{M}^{(K)}\cdots\mathbf{M}^{(k+1)}\mathbf{V}^{(k)}\text{diag}\left[\sigma_{k,1}, \sigma_{k,2}, \ldots, \sigma_{k,D}\right]\mathbf{n}.$$

The singular value decomposition of $\mathbf{M}^{(K)}\cdots\mathbf{M}^{(k+1)}\mathbf{V}^{(k)}\text{diag}\left[\sigma_{k,1}, \sigma_{k,2}, \ldots, \sigma_{k,D}\right]$ may be assumed to yield $\tilde{\mathbf{V}}_n^{(k)}\tilde{\Lambda}_n^{(k)}\tilde{\mathbf{U}}_n^{(k)}$, with output noise correlation matrix equal to $\tilde{\mathbf{V}}_n^{(k)}(\tilde{\Lambda}_n^{(k)})^2\tilde{\mathbf{V}}_n^{(k)*}$, which is of the same form as Eq. (16), with $\tilde{\mathbf{V}}_n^{(k)}$ independent of $\mathbf{V}^{(K)}$.

When the total number of noise sources is very large, by the law of large numbers [49, Sec. 3-3], the overall noise correlation matrix converges to a form similar to Eq. (17). The law of large numbers is applicable provided the number of noise sources is large and the $\tilde{\mathbf{V}}_n^{(k)}$ for each *k* indexing a noise source are independent, random unitary matrices.[6] Based on the law of large numbers, at the fiber output, the noises in the different modes are i.i.d. Ideally, the receiver uses $\mathbf{V}^{(t)*}$ to diagonalize the channel, as in Eq. (6). After diagonalization, provided the number of noise sources is large and the $\mathbf{V}^{(t)*}\tilde{\mathbf{V}}_n^{(k)}$ for each *k* indexing a noise source are independent, random unitary matrices, the law of large numbers is applicable, and the noises in the different diagonal spatial channels are i.i.d. Because of the relationship between $\mathbf{V}^{(t)}$, $\tilde{\mathbf{V}}_n^{(k)}$ and the individual $\mathbf{U}^{(k)}$ and $\mathbf{V}^{(k)}$, $\mathbf{V}^{(t)*}\tilde{\mathbf{V}}_n^{(k)}$ are random unitary matrices, except in special cases.

We have studied MDL statistics only for a single frequency. MDL is frequency-dependent with a certain coherence bandwidth, which is expected to depend on the modal dispersion in Eq. (3). MDL should be statistically independent for frequencies whose separation is much larger than a certain coherence bandwidth, which should be of the same order as $1/\sigma_{gd}$, where $\sigma_{gd}$ is the STD of the group delay [7]. If the overall bandwidth of a signal is much larger than the coherence bandwidth, due to frequency diversity, the overall outage capacity should approach the average capacity (shown in Figs. 6 and 7 for cases without or with CSI), instead of the single-frequency outage capacities shown in Fig. 8. As explained earlier, modal dispersion does not fundamentally degrade performance, but does affect receiver complexity.

## 6. Conclusions

Two main propositions have been verified here numerically. The MDL in MMF is statistically the same as the eigenvalue distribution of a zero-trace Gaussian unitary ensemble in the small-MDL regime. The STD of the overall MDL may be approximated by Eq. (1) over a wide range of MDL values. Depending on the square-root of the accumulated MDL variance $\xi = \sqrt{K}\sigma_g$, both propositions are valid up to $\xi = 10$ dB for two-mode fibers and up to $\xi = 15$ dB for fibers with many modes. This range of validity corresponds to a maximum MDL difference up to 23.4 dB in two-mode fibers and nearly 80 dB in fibers with many modes, far larger than values likely to be acceptable for practical long-haul systems.

---

[6] The law of large numbers is concerned with averages. The overall noise correlation matrix is a summation over the correlation matrices corresponding to all the independent noise sources. The normalized cross-correlation of the noise is characterized by ratios between its off-diagonal elements and its diagonal elements, which are variances proportional to the number of noise sources. Thus, the normalized cross-correlation is implicitly an "average", to which the law of large numbers is applicable.

The channel capacity of MMF has been calculated based on the proposition that the MDL in MMF has the same statistical properties as a zero-trace Gaussian unitary ensemble. Both average and outage capacities, with and without CSI, match results of brute-force simulation in the small-MDL regime.

**Acknowledgments**

The research of JMK was supported in part by National Science Foundation Grant Number ECCS-1101905 and Corning, Inc.

**Appendix: Generation of Random Matrices**

The generation of the unitary matrices $\mathbf{U}^{(k)}$ and $\mathbf{V}^{(k)}$ may employ the Gram-Schmidt process [57, Sec. 5.2.7]. *D* random complex Gaussian vectors are generated as initial vectors, and the Gram-Schmidt process is used to obtain *D* orthonormal vectors. Those *D* orthonormal vectors are combined to form a unitary matrix.

A zero-trace Hermitian Gaussian matrix may be generated by the following procedure. Given a random complex Gaussian matrix $\mathbf{A}$, $\mathbf{A} + \mathbf{A}^*$ is a Hermitian-Gaussian matrix [58], but does not necessarily have zero trace. A zero-trace matrix is obtained by replacing the diagonal elements of $\mathbf{A} + \mathbf{A}^*$ by *D* real Gaussian-distributed random numbers that sum to zero. Using the Gram-Schmidt process with the first vector as [1, 1, …, 1] and the remaining $D - 1$ vectors initialized by real Gaussian vectors, $D - 1$ orthonormal vectors are obtained in which the vector elements sum to zero. Those $D - 1$ orthonormal vectors may be combined to form a $D \times (D-1)$ real matrix $\mathbf{Q}$. Multiplication of $\mathbf{Q}$ with a vector of $D-1$ Gaussian random numbers gives *D* Gaussian-distributed numbers that sum to zero. Those *D* Gaussian numbers may be scaled to match the variance of the non-diagonal elements of $\mathbf{A} + \mathbf{A}^*$ and used to replace the diagonal elements of $\mathbf{A} + \mathbf{A}^*$, yielding a zero-trace Hermitian Gaussian matrix. Only one instance of the real matrix $\mathbf{Q}$ is required for each dimension *D*.

Instead of using the Gram-Schmidt process, for certain values of *D*, Hadamard matrices can be employed [59, ch. 7]. Hadamard matrices are orthogonal matrices with all matrix elements equal to $\pm 1$, and typically with a first column comprising all ones. In these cases, the real matrix $\mathbf{Q}$ may be obtained by deleting the first column of a Hadamard matrix.

The eigenvalues of a tri-diagonal random matrix can be the same as those of a Gaussian unitary ensemble [58]. Using a procedure similar to that described above, a tri-diagonal matrix may be modified to have zero trace. The eigenvalues of a tri-diagonal matrix can be found with far fewer operations than those of a fully populated matrix [57, Sec. 8.5].